\documentclass{desyproc}
\begin{document}
\title{DARWIN}

\author{{\slshape Marc Schumann for the DARWIN consortium\footnote{The following institutions are members of the DARWIN consortium (status October 2011): 
ETH~Z\"{u}rich, U~Z\"{u}rich (CH); 
Karlsruhe, Mainz, MPIK~Heidelberg, M\"{u}nster (DE); 
Subatech (FR); 
Weizmann Institute of Science (IL);
INFN (IT): Bologna, L'Aquila, LNGS, Milano, Milano Bicocca, Napoli, Padova, Pavia, Perugia, Torino; 
Nikhef (NL);
Associated US Groups: Columbia, Princeton, UCLA, Arizona State.}}\\[1ex]
Physics Institute, University of Z\"urich, CH-8057 Z\"urich, Switzerland
}

\contribID{schumann\_marc}

\desyproc{DESY-PROC-2011-04}
\acronym{Patras 2011} 
\doi  

\maketitle

\begin{abstract}
DARWIN is a design-study for a next-to-next generation experiment to directly detect WIMP dark matter in a detector based on a liquid xenon/liquid argon two-phase time projection chamber. This article describes the project, its goals and challenges, and presents some of the recent R\&D highlights.
\end{abstract}

\section{Introduction}

Even though the existence of dark matter in the Universe is well established since many years~\cite{ref::pdg2010}, the nature of the dark matter particle remains unknown and has not been detected  convincingly yet. One of the most promising particle candidate is the weakly interacting massive particle (WIMP), which arises naturally in many extensions of the standard model.

\begin{figure}[h!]
\includegraphics[width=0.48\textwidth]{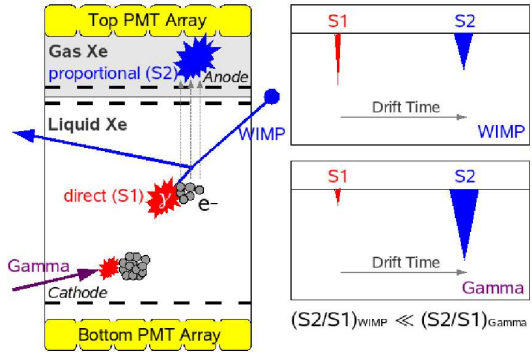}\hspace{0.04\textwidth}
\begin{minipage}[b]{0.48\textwidth}\caption{Principle of a two-phase liquid-noble gas TPC: Interactions generate prompt scintillation light (S1) and ionize the liquid target. The ionization electrons (charge signal, S2) are drifted upwards and detected in the gas volume. The event vertex is reconstructed from the S2~pattern and the electron drift time. The S2/S1 ratio 
is used for signal/background discrimination. In LAr, the S1 pulse-shape provides another powerful discrimination channel. \label{fig::twophase}}
\end{minipage}
\end{figure}

Detectors using the liquid noble gases xenon (LXe)~\cite{ref::xenon, ref::zeplin} and argon (LAr)~\cite{ref::warp} as a WIMP target in a two-phase time projection chamber (TPC), see Fig.~\ref{fig::twophase}, have already delivered very competitive results. They are also among the most promising techniques to build future detectors with a multi-ton target mass and an ultra-low background. These are the crucial requirements in order to be able to detect a significant number of WIMPs to allow for a clear dark matter detection. If the WIMP is found with the current or next generation detectors, it would allow for ``WIMP spectroscopy'' to constrain its properties.

DARWIN (dark matter WIMP search with noble liquids)~\cite{ref::darwin} is an international consortium which brings together experts from the LAr experiments WARP~\cite{ref::warp}, ArDM~\cite{ref::ardm} and DarkSide~\cite{ref::ds} and from the LXe experiment XENON~\cite{ref::xenon}, as well as new particle physics groups, in order to design Europe's next-to-next generation dark matter facility based on LXe and/or LAr. 
DARWIN is supposed to be an ultimate WIMP detector, limited only by irreducible neutrino backgrounds and finite rejection of background events from the bulk.
The outcome of the design study, which is funded by ASPERA since late 2009, is a proposal for an experiment. 
The details, including the decision whether the facility will be based on LXe and/or LAr and at which target mass, are not decided yet and are part of the study. The idea behind the combination of LXe and LAr in one experiment is to use synergies between both communities and to exploit some complementarity of the different targets~\cite{ref::darwin,ref::pato,ref::ucla}.

\section{Goals, Challenges and Progress}\label{sec::goals}

R\&D within the DARWIN project is organized in seven work packages (plus one for management): Detector Infrastructure, Light Readout and Light Response, Charge Readout, Electronics and Trigger, Underground Site, Shields and Background, Material Selection and Screening (plus measurement of minute radioactive traces), and Science Impact. A biased selection of intermediate results is presented in the remaining part of this section.

\begin{figure}[b!]
\centering
\includegraphics[width=0.98\textwidth]{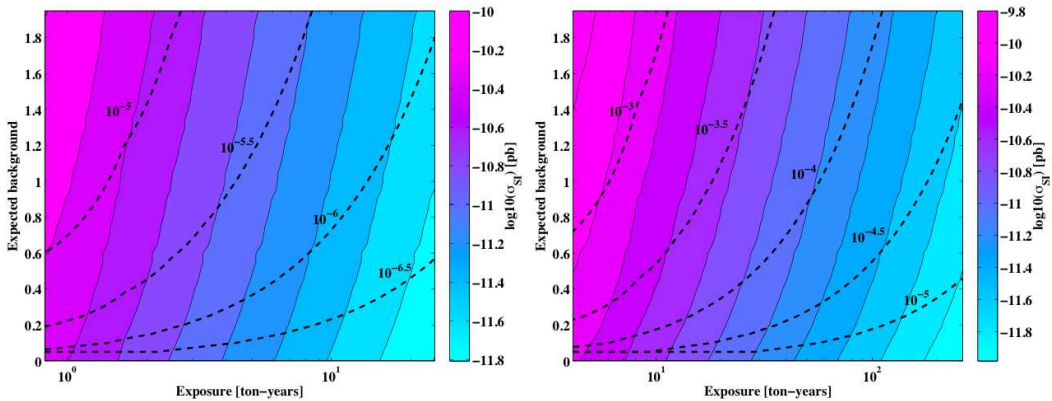}\vspace{-3mm}
\caption{Background requirements for a multi-ton LXe (left) and LAr TPC (right). Shown is the number of background events vs exposure (the dashed lines show the background in dru = evts$\times$keV$^{-1}\times$kg$^{-1}\times$day$^{-1}$). The $z$-axis gives the sensitivity to spin-independent WIMP nucleon cross sections $\sigma$ (in pb). The state-of-the-art values for background, acceptance, threshold (10~keVr/30~keVr for LXe/LAr), and background rejection (99.5\%/$3\times10^{-7}$) are used in the study. To reach $\sigma \sim 10^{-11}$~pb=$10^{-47}$~cm$^2$, a background level of $10^{-5.5}$--$10^{-6}$~dru is required for LXe in a 5~ton$\times$year exposure and about $10^{-4}$ for LAr in a 40~ton$\times$year exposure. \label{fig::bg}}
\end{figure}

{\bf Background: } A next-to-next generation dark matter experiment must aim at a sensitivity to spin-independent WIMP nucleon cross sections around $10^{-48}$~cm$^2$, well below the goal of the next generation experiments such as XENON1T, DarkSide, and SuperCDMS. Besides the target mass, for which we assume as benchmark cases at least 5~tons of LXe and $\sim$10~tons of LAr after fiducial volume cuts, one of the most crucial requirements is an extremely low background. Radioactivity from the detector itself can be reduced very efficiently by a fiducial volume cut and background from neutrons can be tackled by large water- or scintillator-based shields and a muon veto, as well as exploiting the possibility to detect and reject multiple-scatter interactions with the TPC. The remaining background is from intrinsic radioactive contamination in the target itself, which is mainly $^{85}$Kr and $^{222}$Rn for LXe and $^{39}$Ar (and to a lower extent also $^{85}$Kr and $^{222}$Rn) in LAr. Fig.~\ref{fig::bg} shows the impact of background on DARWIN's sensitivity. It becomes clear that the background reduction and discrimination standards, which have been achieved by the current generation of experiments, are not yet sufficient for DARWIN. The Kr~concentration in LXe, for example, must be reduced to below 1~ppt, and the background rejection by the pulse-shape analysis method in LAr must be improved by an order of magnitude.

Progress is also made regarding the development of the the tools to measure minute traces of radioactivity in liquid noble gases at several member institutions (MPIK, M\"{u}nster, Columbia) using different methods such as mass spectroscopy, gas chromatography, atom trapping etc.

{\bf Light response of LXe/LAr to low energy interactions: } Several experiments are ongoing in order to study the response of the noble liquid targets to low energy nuclear recoils (from neutron or WIMP interactions) or electronic recoils (from $\gamma$-interactions). The Columbia group has recently published a new measurement of the relative scintillation efficiency ${\cal L}_\textnormal{\footnotesize eff}$ in LXe~\cite{ref::leff} which clearly shows non-zero values down to the lowest measured data-point of 3~keVr (keV nuclear recoil energy). A similar measurement for LAr is currently ongoing at CERN/U~Z\"{u}rich~\cite{ref::regenfus}. A measurement of the LXe response to electronic recoils down to 2.5~keV is currently performed at U~Z\"{u}rich~\cite{ref::manalaysay} and also indicates non-zero values.

\begin{figure}[b!]
\begin{minipage}[]{0.48\textwidth}
\centering
\includegraphics[width=0.9\textwidth]{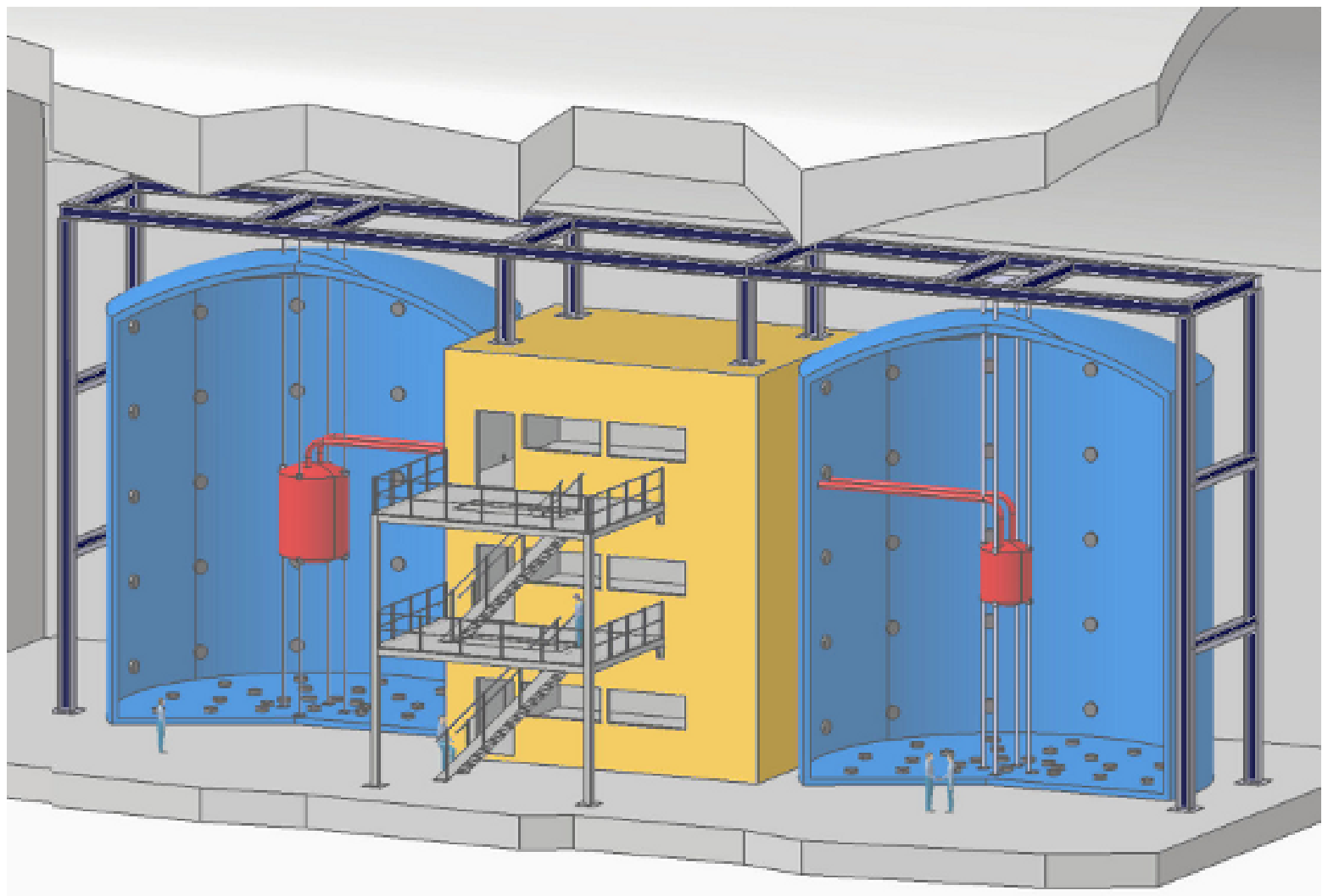}
\caption{Illustration of a possible layout of the DARWIN facility: A 10~ton LXe detector (5~ton fiducial mass) and a 20~ton LAr detetector (10~ton fiducial mass). In the final realization, the two detectors do not necessarily need to be placed directly next to each other.\label{fig::layout}}
\end{minipage}
\hfill
\begin{minipage}[]{0.48\textwidth} 
\centering
\vspace{-0.15cm}
\includegraphics[width=\textwidth]{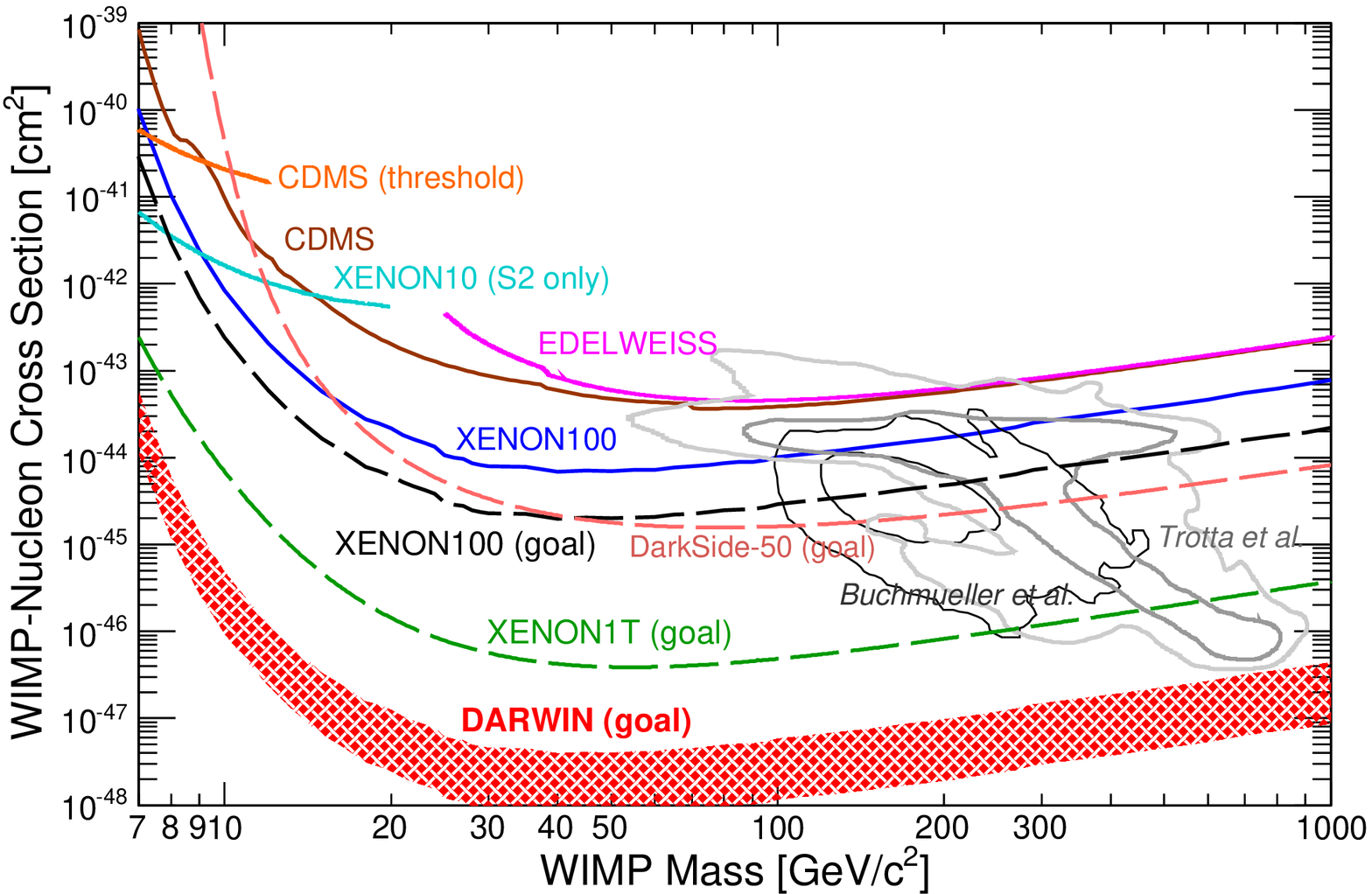}
\caption{DARWIN's sensitivity goal is in the low $10^{-48}$~cm$^2$ range for spin-independent WIMP nucleon cross sections, more than one order of magnitude below the goals of the next-generation experiments. Solid curves indicate published results.}
\label{fig::limit}
\end{minipage}
\end{figure}

{\bf Light and charge readout: } Two main avenues are currently pursued to detect the prompt scintillation light from the interactions. One involves the QUPID, a novel hybrid photodetector with very low intrinsic radioactivity~\cite{ref::qupid} which has been developed by UCLA for use in LXe and LAr. Another approach is the commercial 3''~photomultiplier tube R11410/R11065 (LXe/LAr) from Hamamatsu, which is extensively tested at several member institutions (see e.g.~\cite{ref::r11065}).

Alternatives to the currently employed charge readout via a secondary scintillation process are interesting when large areas have to be covered and single electron sensitivity and fine spatial granularity are desirable. 
R\&D towards this goal is performed at several institutions, focussing mainly on large cryogenic LEMs/ThickGEMs/Micromegas for noble liquids~\cite{ref::lem} (ETHZ), on gaseous PMTs (GPMs) without dead zones~\cite{ref::gpm} (Subatech/WIS), and on CMOS pixel detectors coupled to electron multipliers (GridPix~\cite{ref::gridpix}, Nikhef).

{\bf Science impact: } Several questions have been addressed within the context of this work package. One is the optimum scaling of the LXe/LAr target mass in order to reach a similar sensitivity (to a 100~GeV/$c^2$ WIMP). It turns out that the optimum ratio would be 1:7, assuming a lower threshold of 10~keVr and 30~keVr for LXe and LAr, respectively. For a target mass of 5~tons of LXe, this would mean a LAr target of 35~tons, challenging in terms of threshold and light detection (necessary for background reduction by the S1 pulse shape analysis). Another topic is the complementarity of the targets, this has been studied in a dedicated publication~\cite{ref::pato}.

\section{The Future}

The outcome of the DARWIN project will be a design report for a multi-ton LXe/LAr dark matter detection experiment, therefore the details are not yet finalized. Nevertheless, Fig.~\ref{fig::layout} illustrates how the future DARWIN facility might look like. The Figure shows two separate large water shields -- instrumented as active \u{C}erenkov muon veto -- housing a 10~ton LXe detector and a 20~ton LAr detector with 5~tons and 10~tons fiducial mass, respectively. If the background requirements as mentioned in Sect.~\ref{sec::goals} are achieved, this facility will be able to probe spin-independent WIMP nucleon cross sections around $10^{-48}$~cm$^2$. This is illustrated in Fig.~\ref{fig::limit}, together with results which have already been achieved (solid lines) and the goals of upcoming experiments (dashed). Also indicated are expectations from some supersymmetric models (gray contours). These predict the WIMP to be in a region of the parameter space which will be completely covered by DARWIN.

\begin{footnotesize}

\end{footnotesize}


\begin{thebibliography}{99}
\bibitem{ref::pdg2010}
K.~Nakamura et al.~(Particle Data Group), \emph{J. Phys.} {\bf G 37}, 075021 (2010) and references therein.

\bibitem{ref::xenon}
J.~Angle et al.~(XENON10), \emph{Phys.~Rev.~Lett.~}{\bf 100}, 021303 (2008); \\
E.~Aprile et al.~(XENON100), \emph{Phys. Rev. Lett.~}{\bf 107}, 131302 (2011).

\bibitem{ref::zeplin}
D.Yu~Akimov et al.,~(ZEPLIN-III), {\tt arXiv:1110.4769} (2011).

\bibitem{ref::warp}
R.~Brunetti et al. (WARP), Astropart.~Phys.~{\bf 28}, 495-507 (2008).

\bibitem{ref::darwin} 
L.~Baudis (DARWIN), PoS(IDM2010)122 (2010). {\tt http://darwin.physik.uzh.ch}

\bibitem{ref::ardm}
A.~Rubbia et al.~(ArDM), J.~Phys.~Conf.~Ser.~{\bf 39}, 129 (2006).

\bibitem{ref::ds} 
A.~Wright (DarkSide), {\tt arXiv:1109.2979} (2011).

\bibitem{ref::pato}
M.~Pato et al., Phys.~Rev.~{\bf D83}, 083505 (2011).

\bibitem{ref::ucla}
K.~Arisaka et al., {\tt arXiv:1107.1295} (2011).

\bibitem{ref::leff}
G.~Plante et al., Phys. Rev. {\bf C84}, 045805 (2011).

\bibitem{ref::regenfus}
C.~Regenfus, talk presented at TAUP2011 (2011).

\bibitem{ref::manalaysay}
A.~Manalaysay, talk presented at TAUP2011 (2011).

\bibitem{ref::qupid}
A.~Teymourian et al., Nucl.~Instr.~Meth.~A {\bf 654}, 184 (2011).

\bibitem{ref::r11065}
R.~Acciarri et al.~(WARP), {\tt arXiv:1108.5584} (2011).

\bibitem{ref::lem}
A.~Badertscher et al, Nucl.~Instr.~Meth.~A {\bf 641}, 48 (2011).

\bibitem{ref::gpm}
S.~Duval et al., {\tt arXiv:1110.6053} (2011).

\bibitem{ref::gridpix}
V.~Blanco Carballo et al., JINST {\bf 5}, P02002 (2010).

\end{thebibliography}
\end{document}